# A 36-Element Solution To Schneiders' Pyramid Hex‑Meshing Problem And A Parity-Changing Template For Hex-Mesh Revision

Shang Xiang and Jianfei Liu[*]

Peking University

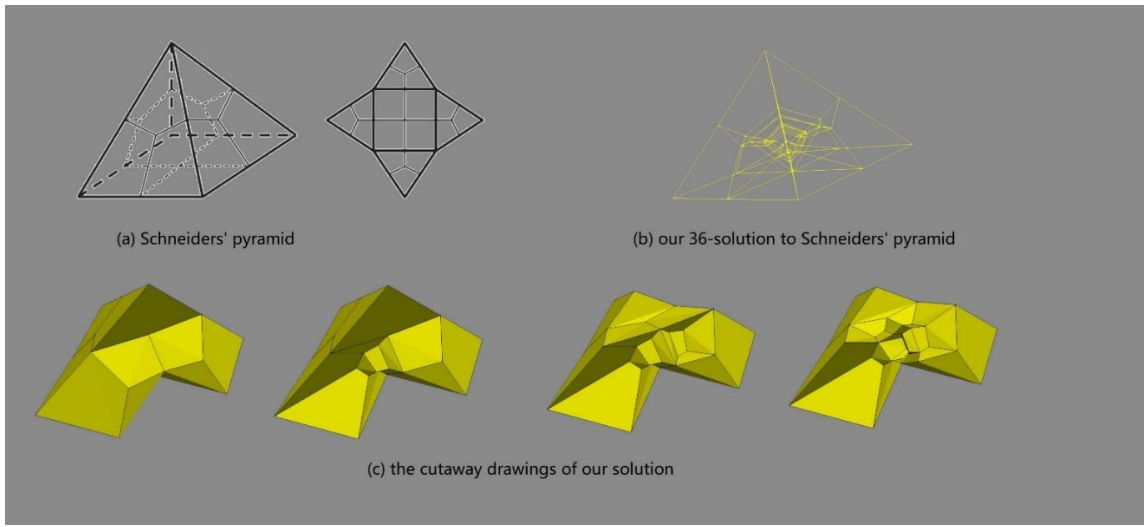

**Figure 1:** *Shcneiders' problem and our 36-element solution*

## ABSTRACT

In this paper, we present a solution that uses the least number of hexahedra to build a pyramid, which is the key block required for one type of automatic hex-meshing method to be successful.

When the initial result of a hex-meshing program is not appropriate for specific applications, some templates are used for revision. The templates reported thus far are parity-preserving, which means that the parity of the number of hexahedra in a mesh is unchanged after a revision following the templates. We present a parity-changing template that makes the template set integral and more effective.

These two findings are obtained by a program that we developed for this study, which is a tool for researchers to observe the characteristics of small hexahedral packings.

[*]To whom correspondence should be addressed. Email: liujianfei@pku.edu.cn.

**ACM Reference format:**

## 1 INTRODUCTION

Hex meshes are urgently needed in various engineering areas. Due to the conforming requirement shown in figure 2, hex meshing is extremely challenging, and it costs days or even weeks to build a moderate complexity mesh for an experienced engineer. For a half century, researchers have attempted to develop methods to automatically construct a hex mesh [Blacker 2001; Owen 1998; Tautges 2001; Fang et al 2016; Li et al 2011; Nieser et al 2011]. However, there is still a long way to make these methods practical.

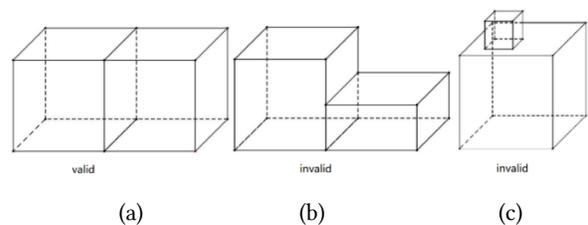

**Figure 2：** *The conforming requirement of a hex mesh: if two hexahedra meet at a face, they should meet at the whole face and not part of it. Case (a) is conforming; cases (b) and (c) are non-conforming.*

A compromise method of automatic hex meshing is to create a hex-dominant mesh first and then transform it to an all-hex mesh. A hex-dominant mesh needs two other types of elements, a pyramid and tetrahedron, to completely fill a space.

There are different types of hex-dominant meshing methods [Stephenson et al 1992; Tristan et al 2014; Sokolov et al 2016; Xifeng Gao et al 2017], yet we focus on the advancing front approach for discussion here since the transformation stage is the same if all-hex meshes are needed.

For a given 3D space, we place hexahedra around the space boundary first and then toward the interior. Due to the conforming requirement, one usually cannot fill the space completely by advancing the front, and a void will be left in the interior, as figure 3 illustrates in the 2D situation. A hex-dominant meshing program will then use the other two types of elements to fill the void. The void is enclosed by quadrilaterals, and from each of the quadrilaterals, a pyramid will grow, as shown in figure 4, which turns the boundary faces of the void into triangles. The void will then be filled with tetrahedrons, and since it is now enclosed by triangles, it is guaranteed to be fulfilled [Bern 1993].

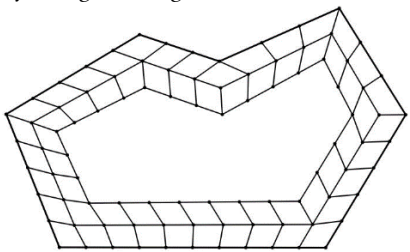

**Figure 3**: *A void is usually left after laying hexahedra in an advancing front manner, as shown in the 2D situation.*

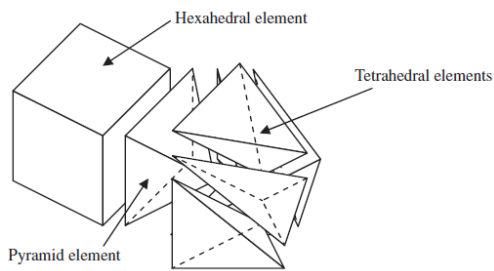

**Figure 4**: *From [Yamakawa et al 2010]. Growing a pyramid from each quadrilateral on the boundary of the void turns the boundary faces into triangles, which guarantees the successive tetrahedral filling.*

The process mentioned above finishes to fill a space with a hex-dominant mesh; one can then transform it to an all-hex mesh by subdividing its elements. Each hexahedron is divided into 8 smaller ones, as shown in figure 5(a), and each tetrahedron is divided into four hexahedra, as shown in figure 5(b). After the subdivision of all the tetrahedrons and hexahedra, the surface of each pyramid will be composed of 16 quadrilaterals, as shown in figure 1(a).

How can a pyramid be subdivided into hexahedra that conform to these 16 quadrilaterals? We now face Schneiders' problem.

**Schneiders' problem** [Schneiders 1996]: creating a hex mesh of a pyramid that conforms to the surface subdivision shown in figure 1(a).

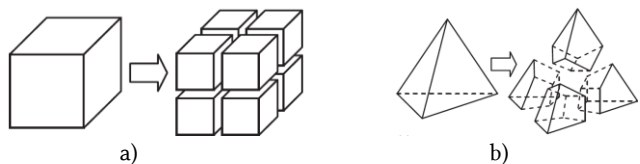

**Figure 5**: *Subdivision patterns for previous hexahedral and tetrahedral elements: (a) dividing one previous hexahedron into 8 smaller ones, and (b) one previous tetrahedron into 4 hexahedra. From [Yamakawa et al 2010].*

Unlike its first appearance, the problem is very difficult. One must grow hexahedra based on the 16 given quadrilaterals toward the interior and make the pyramid to be filled completely.

Obviously, the solution to Schneiders' problem is a key step for building a hex mesh from a hex-dominant mesh.

To solve the problem, we developed a program, which is actually a tool to study small hexahedra packings and, by using it, we have gotten two findings including the solution to the Schneiders' problem.

**Contributions**: We give a 36-element solution to Schneiders' problem, which is the least element solution to the problem. We report a parity-changing template, which is being expected by the community for years. We developed a program, which can be used to study the characteristics of small hexahedra packings.

**Previous works**: Yamakawa et al. [2002] presented a solution for the pyramid problem, which has 118 elements. The same authors gave an improved solution in [Yamakawa et al 2010], which has 88 elements, as shown in figure 6.

The 88 solution is currently the best result; however, it still has the disadvantage of having too many hexahedra because, for a common engineering problem, one will obtain a rather large number of pyramid elements in a hex-dominant mesh. Multiplying by 88 times will produce an enormous number of hexahedra in the final all-hex mesh, which has the consequence of heavily degrading the performance of its subsequent engineering applications.

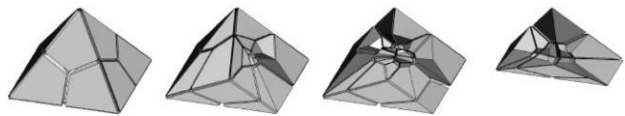

**Figure 6**: *Cutaway drawings of the 88 elements solution published in [Yamakawa et al 2010].*

It is notoriously difficult to improve topology connections of a hex-mesh. Bern et al [2002] studied the basic operations to flip hex-meshes. They claimed a parity-changing template should be existed and it is important theoretically. Yet no parity-changing template has been given so far.

The rest of the articles are organized as follows: section 2 gives our method to solve the Schneider's problem. A program is developed for the purpose. Section 3 uses the program to solve another problem, building a parity-changing template. Section 4 is the discussion and conclusions. The appendix lists the 36-element solution and the parity-changing template.

## 2 AN OPTIMAL SOLUTION GIVEN BY A SYSTEMATIC SEARCHING METHOD

Unlike the 88 and 118 solutions, which are created based on the human imagination, our answer is produced by systematically searching by a computer program.

We first develop a program that adds hexahedra one by one to a hexahedral packing and, for a certain number of hexahedra, checks all the possibilities of packing them together and monitors the outer surface of the packing until it shows a pattern similar to that in figure 1(a).

Let us show the process more clearly. At the starting point, we have only the first hexahedron in the packing, and thus, there is a surface pattern with 6 quadrilaterals, as shown in figure 7(a). Adding the second hexahedron, we then obtain the pattern shown in figure 7(b), with 10 quadrilaterals. Adding the third hexahedron, we can obtain 3 patterns as in figure 7(c) (d) and (e) with 14, 14 and 12 quadrilaterals, respectively. Growing one more hexahedron from figure 7(c), we obtain the 5 situations shown in figure 8(a) through 8(e). Note that the geometry is ignored completely at the current stage.

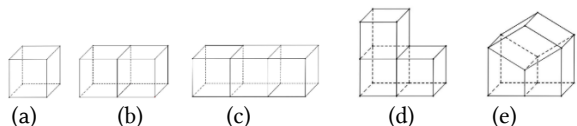

(a)   (b)   (c)   (d)   (e)

**Figure 7**: *The surface patterns of the hexahedral packing with one through three hexahedra*

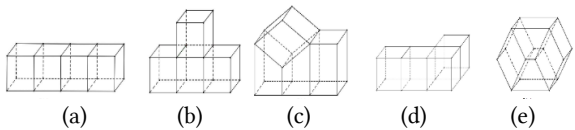

(a)   (b)   (c)   (d)   (e)

**Figure 8**: *The surface pattern of growing one more hexahedron from figure 7(c)*

Continuing to add hexahedra and revise the surface of the packing, the same surface patterns could appear more than once during the process. We keep only the optimal packing, specifically, the packing with the smallest number of hexahedra for each specific surface pattern.

A revision of the surface occurs locally when a new hexahedron is added. Figure 9 demonstrates more clearly the revision from figure 7(b) to figure 7(e), which affects the part colored in red.

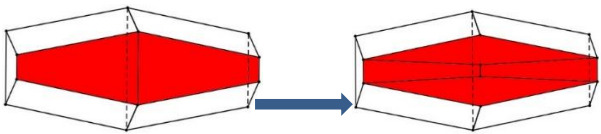

**Figure 9**: *Revision of the packing surface occurs locally*

There are 8 ways in total shown in figure 10, and in one, the surface changes; the figure represents growing a new hexahedron based on one through five quadrilaterals on the surface.

Each subfigure in figure 10 should be read from two opposite directions. Let us take figure 10(a) for an explanation. Reading from left to right results in growing a hexahedron based on 1 quadrilateral, and after the growing, the quadrilateral is replaced by five new ones. Reading in the opposite direction results in growing a hexahedron based on 5 neighboring quadrilaterals, and they will disappear and be replaced by one new quadrilateral on the surface.

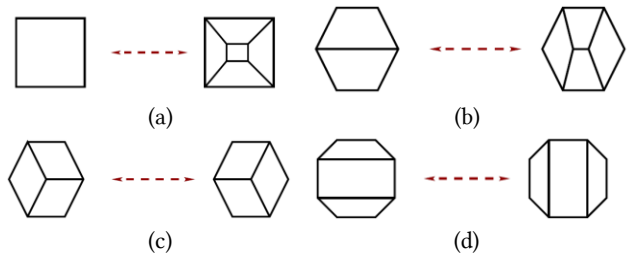

(a)                    (b)

(c)                    (d)

**Figure 10**: *The 8 ways; in one, the surface will change when a new hexahedron is added in a packing.*

The packing process stops when it reaches the pattern shown in figure 1(a), and thus, we obtain the optimal solution to Schneiders' problem.

Based on the idea above, we develop a program, and after approximately 20 hours of running in a PC with i5-4590, it gives the optimal solution for Schneiders' problem.

The solution has 36 hexahedra, which is much less than 88 and thus decreases dramatically the element numbers in a final all-hex mesh.

Each hexahedron has 8 vertices, and the new solution has 51 vertices in total, with 18 on the surface of the pyramid and 33 inside. We list its hex-vertex relationships in table 1 of the appendix.

The geometry is ignored totally in our searching process. Now, it is time to give the vertices geometric positions. The boundary vertices are prescribed. Using the tools for mesh smoothing [Livesu et al 2015], we obtain the positions for the interior vertices. All of the coordinates of the vertices are listed in table 2 of the appendix. The wireframe and cutaway drawings of the solution are shown in figure 1(b) and (c) respectively.

## 3 PARITY-CHANGING TEMPLATE

The program is in fact a tool to research local patterns for hex meshing. We now turn to searching a parity-changing template. A template is a pair of hex meshes with the same quadrilateral surface pattern. If the two hex meshes of a template have different parities, with one having an odd number of elements and the other having an even number of elements, it is then a parity-changing template. Degraded to the 2D case, a parity-changing template for a quadrilateral mesh can be easily given, as in figure 11. Thus far, all of the hex mesh templates reported are parity-preserving and parity-changing templates that exist only as a theoretical imagination [Bern et al 2002].

The mesh quality is important in applications, and there are various measures of the mesh quality [Livesu et al 2015]. The elements subdivided from pyramids and tetrahedrons usually measure badly, which is why we consider that creating meshes in a hex-dominant manner is a trade-off.

A parity-changing template is important because it is the missing part for performing transformations on hex meshes following templates; such a transformation is a way to improve the mesh quality and has a central position in hex-meshing research.

We now present the parity-changing template obtained by our program. The odd mesh in the template has 17 elements, as listed in table 3 of the appendix, and the even mesh has 18 elements, as listed in table 4 of the appendix; they occupy the same space enclosed by 34 quadrilaterals.

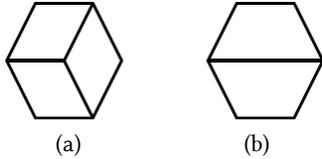

(a)　　　　　(b)

**Figure 11**：*Parity-changing template in quadrilateral meshes. Two different meshes for the same 2D space: (a) has an odd number of quadrilaterals, and (b) has an even number of quadrilaterals*

## 4 CONCLUSIONS AND DISCUSSIONS

Schneiders' problem is the hinge to automatically create a hex mesh by making a hex-dominant mesh first and then transforming it. In this paper, a program is developed to solve the problem, and it successfully obtains the optimal solution. In addition, a parity-changing template, which has been anticipated by researchers for more than a decade, is produced by the program. We believe that it can be used more generally, to obtain local hex mesh patterns with various requirements. Further advances are expected, such as more basic transformations beyond those listed in [Hecht et al 2012].

## ACKNOWLEDGMENTS

.

## A  OUR SOLUTION TO SCHNEIDERS' PROBLEM AND THE PARITY-CHANGING TEMPLATE

Table 1 shows 8 vertices for each of the 36 hexahedra. The order of vertices at a hexahedron is shown in figure 12. The coordinates for each vertex are listed in table 2.

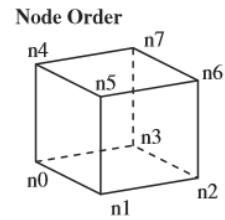

**Figure 12**：*Vertex order for a hexahedron.*

**Table 1:** *Hex-vertex relationships for the 36 elements solution.*

| N0 | N1 | N2 | N3 | N4 | N5 | N6 | N7 |
|----|----|----|----|----|----|----|----|
| 7  | 46 | 48 | 8  | 3  | 50 | 49 | 2  |
| 3  | 50 | 49 | 2  | 18 | 42 | 44 | 6  |
| 50 | 3  | 7  | 46 | 42 | 18 | 19 | 41 |
| 29 | 35 | 36 | 32 | 48 | 46 | 50 | 49 |
| 41 | 42 | 40 | 38 | 46 | 50 | 36 | 35 |
| 40 | 42 | 44 | 31 | 36 | 50 | 49 | 32 |
| 34 | 30 | 47 | 45 | 35 | 29 | 48 | 46 |
| 29 | 48 | 49 | 32 | 30 | 47 | 44 | 31 |
| 48 | 49 | 44 | 47 | 8  | 2  | 6  | 12 |
| 46 | 48 | 47 | 45 | 7  | 8  | 12 | 14 |
| 14 | 45 | 43 | 20 | 12 | 47 | 44 | 6  |
| 34 | 30 | 31 | 33 | 45 | 47 | 44 | 43 |
| 45 | 34 | 35 | 46 | 43 | 33 | 38 | 41 |
| 14 | 45 | 46 | 7  | 20 | 43 | 41 | 19 |
| 42 | 44 | 6  | 18 | 40 | 31 | 23 | 39 |
| 33 | 31 | 23 | 25 | 43 | 44 | 6  | 20 |
| 6  | 18 | 39 | 23 | 4  | 5  | 28 | 24 |
| 23 | 39 | 40 | 31 | 24 | 28 | 36 | 32 |
| 38 | 33 | 25 | 37 | 41 | 43 | 20 | 19 |
| 19 | 41 | 42 | 18 | 37 | 38 | 40 | 39 |
| 28 | 36 | 35 | 27 | 39 | 40 | 38 | 37 |
| 37 | 39 | 18 | 19 | 27 | 28 | 5 | 10 |
| 33 | 38 | 37 | 25 | 34 | 35 | 27 | 26 |
| 10 | 27 | 26 | 16 | 19 | 37 | 25 | 20 |
| 28 | 36 | 32 | 24 | 27 | 35 | 29 | 21 |
| 21 | 29 | 30 | 22 | 27 | 35 | 34 | 26 |
| 25 | 33 | 34 | 26 | 23 | 31 | 30 | 22 |
| 29 | 30 | 31 | 32 | 21 | 22 | 23 | 24 |
| 21 | 27 | 10 | 11 | 24 | 28 | 5 | 4 |
| 26 | 27 | 10 | 16 | 22 | 21 | 11 | 13 |
| 6 | 20 | 16 | 13 | 23 | 25 | 26 | 22 |
| 21 | 22 | 23 | 24 | 11 | 13 | 6 | 4 |
| 6 | 20 | 14 | 12 | 13 | 16 | 15 | 17 |
| 14 | 20 | 19 | 7 | 15 | 16 | 10 | 9 |
| 4 | 5 | 0 | 1 | 6 | 18 | 3 | 2 |
| 0 | 5 | 10 | 9 | 3 | 18 | 19 | 7 |

**Table 2:** *Vertex coordinates for the 36 elements solution.*

| vertex | X | Y | Z |
|---|---|---|---|
| 0 | -0.5 | 0.707107 | 0.5 |
| 1 | -1 | 0 | 1 |
| 2 | 0 | 0 | 1 |
| 3 | 4.96E-23 | 0.471405 | 0.666667 |
| 4 | -1 | 0 | 0 |
| 5 | -0.66667 | 0.471405 | 4.96E-23 |
| 6 | 0.009186 | 0 | 0.000686 |
| 7 | 0.5 | 0.707107 | 0.5 |
| 8 | 1 | 0 | 1 |
| 9 | 0 | 1.41421 | 0 |
| 10 | -0.5 | 0.707107 | -0.5 |
| 11 | -1.01829 | 0.003314 | -1.01895 |
| 12 | 1 | 0 | 0 |
| 13 | -0.04976 | 0.018442 | -0.97388 |
| 14 | 0.686509 | 0.443344 | -0.02596 |
| 15 | 0.5 | 0.707107 | -0.5 |
| 16 | 0 | 0.471405 | -0.66667 |
| 17 | 1 | 0 | -1 |
| 18 | -0.27572 | 0.440414 | 0.28359 |
| 19 | 0.005334 | 0.668948 | 0.002122 |
| 20 | 0.294934 | 0.431786 | -0.28917 |
| 21 | -0.27656 | 0.310609 | -0.28184 |
| 22 | -0.12508 | 0.243362 | -0.29753 |
| 23 | -0.0729 | 0.127332 | -0.07918 |
| 24 | -0.29061 | 0.24351 | -0.12871 |
| 25 | 0.112166 | 0.397343 | -0.25545 |
| 26 | -0.01936 | 0.385934 | -0.33159 |
| 27 | -0.21122 | 0.519918 | -0.21691 |
| 28 | -0.32232 | 0.388203 | -0.02424 |
| 29 | 0.006997 | 0.372234 | -0.0023 |
| 30 | 0.039898 | 0.320725 | -0.03565 |
| 31 | 0.006754 | 0.207526 | -0.00172 |
| 32 | -0.02553 | 0.320584 | 0.030273 |
| 33 | 0.140113 | 0.388999 | -0.13541 |
| 34 | 0.088801 | 0.384454 | -0.08386 |
| 35 | 0.006978 | 0.464393 | -0.00101 |
| 36 | -0.07356 | 0.386614 | 0.079351 |
| 37 | -0.07487 | 0.569095 | -0.08067 |
| 38 | 0.006819 | 0.525466 | -0.00036 |
| 39 | -0.24413 | 0.401184 | 0.104663 |
| 40 | -0.12418 | 0.392346 | 0.130793 |
| 41 | 0.088173 | 0.568537 | 0.081139 |
| 42 | -0.09648 | 0.400928 | 0.25029 |
| 43 | 0.261227 | 0.395528 | -0.10855 |
| 44 | 0.08623 | 0.127288 | 0.078788 |
| 45 | 0.338392 | 0.382141 | 0.020653 |
| 46 | 0.224885 | 0.518732 | 0.216195 |
| 47 | 0.302155 | 0.243057 | 0.125624 |
| 48 | 0.290595 | 0.308293 | 0.281277 |
| 49 | 0.136166 | 0.243999 | 0.292329 |
| 50 | 0.032446 | 0.387901 | 0.326499 |

Table 3: *Hex-vertex relationships for the odd mesh.*

| N0 | N1 | N2 | N3 | N4 | N5 | N6 | N7 |
|---|---|---|---|---|---|---|---|
| 29 | 13 | 21 | 35 | 19 | 9 | 16 | 30 |
| 24 | 25 | 20 | 32 | 21 | 35 | 29 | 13 |
| 12 | 4 | 32 | 20 | 19 | 9 | 13 | 29 |
| 10 | 11 | 26 | 15 | 9 | 19 | 30 | 16 |
| 25 | 24 | 15 | 26 | 35 | 21 | 16 | 30 |
| 32 | 20 | 25 | 24 | 4 | 12 | 7 | 6 |
| 12 | 4 | 9 | 19 | 7 | 6 | 10 | 11 |
| 17 | 18 | 10 | 11 | 25 | 24 | 15 | 26 |
| 3 | 2 | 18 | 17 | 7 | 6 | 24 | 25 |
| 6 | 24 | 32 | 4 | 2 | 18 | 34 | 1 |
| 33 | 0 | 1 | 34 | 17 | 3 | 2 | 18 |
| 17 | 3 | 7 | 25 | 33 | 0 | 5 | 31 |
| 0 | 33 | 34 | 1 | 5 | 31 | 32 | 4 |
| 7 | 25 | 31 | 5 | 12 | 20 | 14 | 8 |
| 14 | 8 | 9 | 13 | 31 | 5 | 4 | 32 |
| 31 | 32 | 13 | 14 | 33 | 34 | 23 | 22 |
| 34 | 33 | 17 | 18 | 23 | 22 | 28 | 27 |

Table 4: *Hex-vertex relationships for the even mesh.*

| N0 | N1 | N2 | N3 | N4 | N5 | N6 | N7 |
|---|---|---|---|---|---|---|---|
| 35 | 18 | 24 | 29 | 30 | 10 | 15 | 19 |
| 21 | 20 | 25 | 27 | 24 | 29 | 35 | 18 |
| 26 | 6 | 27 | 25 | 30 | 10 | 18 | 35 |
| 9 | 8 | 12 | 16 | 10 | 30 | 19 | 15 |
| 20 | 21 | 16 | 12 | 29 | 24 | 15 | 19 |
| 27 | 25 | 20 | 21 | 6 | 26 | 5 | 4 |
| 26 | 6 | 10 | 30 | 5 | 4 | 9 | 8 |
| 14 | 13 | 9 | 8 | 20 | 21 | 16 | 12 |
| 0 | 1 | 13 | 14 | 5 | 4 | 21 | 20 |
| 4 | 21 | 27 | 6 | 1 | 13 | 23 | 2 |
| 22 | 3 | 2 | 23 | 14 | 0 | 1 | 13 |
| 14 | 0 | 5 | 20 | 22 | 3 | 7 | 28 |
| 3 | 22 | 23 | 2 | 7 | 28 | 27 | 6 |
| 5 | 20 | 28 | 7 | 26 | 25 | 17 | 11 |
| 17 | 11 | 10 | 18 | 28 | 7 | 6 | 27 |
| 19 | 12 | 7 | 11 | 30 | 8 | 5 | 26 |
| 21 | 27 | 23 | 13 | 24 | 18 | 34 | 32 |
| 33 | 31 | 25 | 17 | 22 | 14 | 20 | 28 |